\documentclass[prd,showpacs,floatfix,nofootinbib,amsmath,amssymb]{revtex4}
\usepackage{graphicx,graphics,color,dcolumn,booktabs,bm}
\usepackage{longtable,lscape}
\usepackage{txfonts}
\usepackage{overpic}
\usepackage{amssymb}
\usepackage{indentfirst}
\usepackage{epsfig}
\usepackage{feynmf}   %{feynmp}
\usepackage{epstopdf}   %{feynmp}
\usepackage{slashed}  %for Feynman symbols
\usepackage{cases}
\usepackage{color}
\usepackage{multirow}
\usepackage{afterpage}

\graphicspath{{Figures/}} %
\graphicspath{{../property_of_higher_bottomonia_20170911/}}

\usepackage[colorlinks, citecolor=blue,anchorcolor=red,menucolor=red, linkcolor=red,filecolor=red,runcolor=red,urlcolor=blue,frenchlinks=red]{hyperref}

\begin{document}

\title{
Magic mixing angles for doubly heavy baryons
}

\author{Takayuki Matsuki$^1$}\email{matsuki@tokyo-kasei.ac.jp}
\author{Dian-Yong Chen$^2$}\email{chendy@seu.edu.cn}
\author{Xiang Liu$^{3,4}$}\email{xiangliu@lzu.edu.cn}
\author{Qi-Fang L\"u$^{5,6,7}$}\email{lvqifang@hunnu.edu.cn}

\affiliation{
$^1$Tokyo Kasei University, 1-18-1 Kaga, Itabashi, Tokyo 173-8602, Japan\\
$^2$School of Physics, Southeast University, Nanjing 210094, China\\
$^3$School of Physical Science and Technology, Lanzhou University, Lanzhou 730000, China\\
$^4$Research Center for Hadron and CSR Physics, Lanzhou University and Institute of Modern Physics of CAS, Lanzhou 730000, China\\
$^5$Department of Physics, Hunan Normal University,  Changsha 410081, China\\
$^6$Synergetic Innovation Center for Quantum Effects and Applications (SICQEA), Changsha 410081, China\\
$^7$Key Laboratory of Low-Dimensional Quantum Structures and Quantum Control\\
 of Ministry of Education, Changsha 410081, China\\
}

\begin{abstract}

\iffalse
\begin{flushright}
{2020/April/25}\\
{2020/May/19}
(2021/07/05)
(2021/08/08
\end{flushleft}
\fi
\iffalse
We examine the so-called magic mixing angles both for heavy-light mesons and doubly heavy baryons. It turns out that a magic mixing angle rotates the states with definite $^{2S+1}L_J$ for a heavy-light meson to make them the heavy-quark symmetric states.
\fi

We study the so-called magic mixing angles for doubly heavy baryons. Defining that a magic mixing angle rotates states with definite $^{2S+1}(l_\lambda)_J$ to make them heavy-quark symmetric states, 
we derive the magic mixing angle only in the case $L_\rho=0$ between the heavy quark symmetric states with quantum numbers $\left(J, j_\ell\right)$ and the states with $\left(J, s_q+j_\rho\right)=\left(J, \{^4l_\lambda/^2l_\lambda\}\right)$ for a doubly heavy baryon in the standard $\rho-\lambda$ configuration, where $\vec j_\ell=\vec l_\lambda+\vec s_q$, $\vec s_\rho=\vec s_{Q1}+\vec s_{Q2}$, and $\vec j_\rho=\vec s_\rho+\vec L_\rho$.
We discuss how this mixing angle affects model building for a doubly heavy baryon.

\end{abstract}

\pacs{}

\maketitle

%%%%%%%%%%%%%%%%%%%%%%%%%%%%%%%%%%%%%%%%%%%%%%%%%%%%%%%%%%%%%%%%%%%%%%%%%%%%%%%%
\section{Introduction}\label{intro}

Recent experiments by the LHCb collaboration \cite{Aaij:2017ueg} inspired many theorists to explain properties of doubly heavy baryons given by either $ccq$ or $bbq$ states with a light quark $q$, and to determine their mass spectrum and spin-parity. 
Please refer to pioneering and latest theoretical works on doubly heavy baryons in Refs. \cite{Capstick:1986bm,Yoshida:2015tia,Ebert:2002ig,Lu:2017meb,Chen:2017sbg,Xiao:2017dly,He:2021iwx}.
Among many novel features of a heavy-light system, a ``magic'' mixing angle of heavy-light mesons between $^3P_1$ and $^1P_1$ states is very remarkable in that it is uniquely determined in the heavy quark limit and is used to obtain physical states from nonrelativistic states. 

{
Mixing can in principle occur when two states have the same total angular momentum and parity $J^P$. Phenomenological mixing between two states is normally considered when two masses are close to each other and of course with the same $J^P$. For instance, $S$-$D$ mixing can be considered between $nS$ and $(n-1)D$ states with principal quantum numbers $n$ and $n-1$ because their masses are ordinarily  close to each other. If we assume that states consist of quarks, there is a physical meaning for the mixing angle even if it describes internal structure of a hadron, and it can be experimentally measured or determined because strong decays of mixed states  heavily depend on the mixing angle. We may easily find the states that match these conditions in heavy-light systems because there are degenerate states in the heavy quark limit and there are two states with the same $J^P$, e.g., $^3P_1$ and $^1P_1$ states with $J^P=1^+$. This mixing angle is called a ``magic mixing angle''. As for strong decays of doubly heavy baryons, there is some confusion how to identify heavy quark symmetric states, which will be discussed in the final section, Sec. \ref{summary}.

The term ``magic mixing angle'' probably first appeared in Ref. \cite{Barnes:2002mu}, in which details of mixing angle dependence of strong decays of the kaons have been throughly studied and $K_1(1273)$ and $K_1(1403)$ are assumed to be mixed with the magic mixing angle $\tan\theta=1/\sqrt{2}$ with $s$ quark mass $m_s\to\infty$ limit. Another thing we have to keep in mind is that although states can be expressed in several independent orthogonal bases, it is important to express states in terms of those with definite $^{2S+1}L_J$ because with these quantum numbers one easily performs convolution integrals of initial and final states.}
Now, we address here the problem whether there may occur a ``magic'' mixing angle for a doubly heavy baryon similar to a heavy-light meson in the sense that a magic mixing angle for a doubly heavy baryon again rotates the non-relativistic states with definite $^{2S+1}L_J$ to make them the heavy-quark symmetric (physical) states. This problem is worthwhile addressing because no paper seems to discuss strong decays using a magic mixing angle for doubly heavy baryons. Before studying a baryon in detail, let us recall how a magic mixing angle occurs and/or is determined for a heavy-light meson.

Having a deep insight into a structure of heavy-light mesons, Rosner expressed these mesons with $L=1$ as admixtures of $^3P_1$ and $^1P_1$ states by considering spin-orbit and tensor forces which may resolve degeneracy of these mesons \cite{Rosner:1986} and obtained a ``magic'' mixing angle as $\arctan\left(1/\sqrt{2}\right)=35.3^\circ$. In the present language, neglecting a tensor force or regarding it as a constant in the heavy quark limit, his consideration gives two equations, one for an eigenvalue equation of mass in Eq. (\ref{eq:eigen}), and another for the relation between heavy-quark symmetric states (lhs) and states with definite $^{2S+1}L_J$ (rhs) in Eq. (\ref{eq:rel}):
\begin{eqnarray}
 M_{\rm split}
 \left( {\begin{array}{*{20}{c}}
   {\left| {^1{P_1}} \right\rangle }  \\
   {\left| {^3{P_1}} \right\rangle }  \\
 \end{array} } \right)
 = 
 \left(M_0 - \left<H_{SO}^{\bar Qq}\right> \tilde O\right)
 \left( {\begin{array}{*{20}{c}}
   {\left| {^1{P_1}} \right\rangle }  \\
   {\left| {^3{P_1}} \right\rangle }  \\
 \end{array} } \right), \quad
 \tilde O =
 \left( {\begin{array}{*{20}{c}}
   1 & \sqrt{2}  \\
   \sqrt{2} & 0  \\
 \end{array} } \right),
 \label{eq:eigen}
\end{eqnarray}
where $\tilde O$ is a matrix form of expectation values of an operator $-2\vec L\cdot\vec s_q$. 
Eq. (\ref{eq:eigen}) is diagonalized by an orthogonal matrix with a mixing angle $\theta=35.3^\circ$, which describes heavy-quark symmetric states as
\begin{eqnarray}
\left( {\begin{array}{*{20}{c}}
   {\left| {{J^P=1^ + },j_\ell = {{\frac{1}
{2}} }} \right\rangle }  \\
   {\left| {{J^P=1^ + },j_\ell = {{\frac{3}
{2}} }} \right\rangle }  \\
 \end{array} } \right) 
 = 
 \left( {\begin{array}{*{20}{c}}
   {\cos \theta } & {\sin \theta }  \\
   { - \sin \theta } & {\cos \theta }  \\

 \end{array} } \right)\left( {\begin{array}{*{20}{c}}
   {\left| {^1{P_1}} \right\rangle }  \\
   {\left| {^3{P_1}} \right\rangle }  \\
 \end{array} } \right), \label{eq:rel}
\end{eqnarray}
where kets on lhs are heavy-quark symmetric states with $J^P$ and light quark degrees of freeddom $j_\ell$.
Following Rosner's paper, there appeared a couple of papers \cite{Godfrey:1991,Cahn:2003,Close:2005,Matsuki:1997da,Matsuki:2010zy,Ebert:2009ua} with more explicit spin-orbit and tensor interactions, which, of course, lead to the same conclusion as Rosner for $L=1$ and can extend it to any larger $L$ state.
{\it The reason why we need to rotate some states with definite $^{2S+1}L_J$ using Eq. (\ref{eq:rel}) is because Nature is approximately heavy quark symmetric and hence heavy quark symmetric states are closer to physical states rather than nonrelativistic eigenstates with definite $^{2S+1}L_J$.} On the other hand, we normally and dynamically solve mass spectrum of heavy-light mesons by using the potential model that respects the ordinary $^{2S+1}L_J$ quantum number.

This conclusion of Eq. (\ref{eq:rel}) should have been obtained if we could start from a heavy quark symmetric theory from the outset and obtain heavy quark symmetric eigenstates. Actually, this has been achieved in Ref. \cite{Matsuki:1997da}, in which a heavy-quark symmetric Hamiltonian is obtained and is solved to get mass spectrum of heavy-light mesons. 
The heavy-quark symmetric Hamiltonian derived in Ref. \cite{Matsuki:1997da} is given by
\begin{eqnarray}
  H_{\bar Qq}=\vec p\cdot\vec\alpha_q + \beta_q m+S(r) + V(r),
  \label{eq:HQq}
\end{eqnarray}
where $\vec\alpha_q$ and $\beta_q$ are Dirac matrices for a light quark, a confining potential $S(r)=r/a^2+b$, and a one-gluon exchange potential $V(r)=-4\alpha_s/(3r)$. This can be also obtained by using the Nambu-Bethe-Slalpeter equation \cite{Roberts:1993}. This simple equation is very persuasive because this expresses dynamics of one light quark having an interaction with static color source due to a heavy quark expressed by $V(r)$ together with confining potential $S(r)$ between them. 
%In this case, the obtained equation is nothing but a relation between heavy-quark symmetric states and states with definite $^{2S+1}L_J$ quantum numbers with a magic mixing angle.  
In Ref. \cite{Matsuki:1997da}, the heavy quark symmetric wave function is expressed in terms of those with definite $^{2S+1}L_J$, which is nothing but Eq. (\ref{eq:rel}) with a magic mixing angle $\theta=35.3^\circ$ for $L=1$.
This Hamiltonian has a special good quantum number expressed as \cite{Rose:1961,Matsuki:1997da,Roberts:1993}
\begin{eqnarray}
  K = -\beta_q\left(\vec\Sigma_q\cdot\vec L +1\right), \quad \left[H_{\bar Qq}, K\right]=0,
  \label{eq:K}
\end{eqnarray}
whose nonrelativistic expression is given by $\tilde K=-\vec L\cdot\vec\sigma_q-1$. Since $\tilde O=-\left<\vec L\cdot\vec\sigma_q\right>$ in Eq. (\ref{eq:eigen}), we obtain its expectation value as $\left<\tilde K\right> = \tilde O - 1$. Because eigenvalues of $\tilde O$ are 2 and -1, those of $\left<\tilde K\right>$ are $k=1$ and -2 corresponding to $j_\ell =1/2$ and 3/2, respectively (see Table XI in Ref. \cite{Matsuki:1997da}), which form mixing states with $^3P_1$ and $^1P_1$. Physical meaning of the operator $K$ in Eq. (\ref{eq:K}) has been not known well for a long time, but it becomes clear now. Its expectation value appears in Eq. (\ref{eq:eigen}) as eigenvalues and is used to classify heavy-light mesons, which naturally leads to Eq. (\ref{eq:rel}).

The free kinetic terms shown in Eq. (\ref{eq:HQq}) implicitly involve this relation, which means they do not need the interaction that Rosner introduced, and the relation appears only between angular and spin dependent wave functions. 
Refs. \cite{Godfrey:1991,Cahn:2003,Close:2005,Ebert:2009ua} have included the interactions similar to Eq. (\ref{eq:eigen}) so that they have seen occurrence of magic mixing and hence they could reproduce Eq. (\ref{eq:rel}).
Some of them also found this relation kinematically using $6J$ symbols, but they did not ``derive'' it from their Hamiltonian as a relation like in Refs. \cite{Matsuki:1997da,Matsuki:2010zy}.

Let us consider the case for any $L$. In this case, mixing states are given by those with $k=L$ and $k=-L-1$ corresponding to $j_\ell=L-1/2$ and $L+1/2$. %Since $| k|=j_\ell+1/2$, $J=L$.
Using the $6J$ symbol, extension of Eq. (\ref{eq:rel}) to any $L$ is given by \cite{Matsuki:1997da,Cahn:2003,Matsuki:2010zy}
\begin{eqnarray}
\left( {\begin{array}{*{20}{c}}
   {\left| {{J=L },j_\ell = L+1/2} \right\rangle }  \\ % upper and lower are interchanged
   {\left| {{J=L },j_\ell = L-1/2} \right\rangle }  \\
 \end{array} } \right) 
& =&
 \frac{1}{\sqrt{2L+1}}\left( {\begin{array}{*{20}{c}}
   {\sqrt{L+1} } & { \sqrt{L} }  \\
   { - \sqrt{L} } & { \sqrt{L+1} }  \\

 \end{array} } \right)\left( {\begin{array}{*{20}{c}}
   {\left| {J=L, S=0} \right\rangle }  \\
   {\left| {J=L, S=1} \right\rangle }  \\
 \end{array} } \right) \label{eq:rel2}
\\
&=&
\frac{1}{\sqrt{2L+1}} \left( {\begin{array}{*{20}{c}}
   {\sqrt{L+1} } & { \sqrt{L} }  \\
   { - \sqrt{L} } & { \sqrt{L+1} }  \\

 \end{array} } \right)\left( {\begin{array}{*{20}{c}}
   {\left| {^1{L_L}} \right\rangle }  \\
   {\left| {^3{L_L}} \right\rangle }  \\
 \end{array} } \right), \label{eq:rel3}
\end{eqnarray}
where a magic mixing angle is given by
\[
  \tan\theta_L = \sqrt{\frac{L}{L+1}},
\]
which, of course, gives $\theta_L=\arctan(1/\sqrt{2})=35.3^\circ$ in the case of $L=1$. For readers' reference, we list the values of each quantum number in Table \ref{table2}.
%%%%%%%%%%%%%%%
\begin{table}[htpb]
\centering \caption{Correspondence of quantum numbers, $^{2S+1}L_J$, $J^P$, $j_\ell$, and $k$, in the case of heavy-light mesons. Here $L\ge=1$.\label{table2}}
\begin{tabular}{ccccccc}
\toprule[1pt]
%State        &~& &~&  &~&  \\
$^{2S+1}L_J$  &~& $J^P$ &~& $j_\ell=L\otimes s_q$ &~& $k$\\
\midrule[0.6pt] %
$^1S_0$ &~& $0^-$ &~& 1/2 &~& -1\\
$^3S_1$ &~& $1^-$ &~& 1/2 &~& -1\\
\hline 
$^3P_0$       &~& $0^+$ &~& 1/2 &~&  1\\
$^3P_1/^1P_1$ &~& $1^+$ &~& 1/2 &~&  1\\
\hline 
$^1P_1/^3P_1$ &~& $1^+$ &~& 3/2 &~& -2\\
$^3P_2$       &~& $2^+$ &~& 3/2 &~& -2\\
\hline 
$\vdots$       &~& $\vdots$ &~& $\vdots$ &~& $\vdots$\\
\hline 
$^3L_L/^1L_L$ &~& $L^{(-)^{L+1}}$ &~& $L-1/2$ &~&  $L$\\
$^1L_L/^3L_L$ &~& $L^{(-)^{L+1}}$ &~& $L+1/2$ &~& $-L-1$\\
\bottomrule[1pt]
\end{tabular}
\end{table}
%%%%%%%%%%%%%%%%%%%%%%%%%%%%
What we would like to do in this paper is to study a possibility whether the relation corresponding to Eq.~(\ref{eq:rel}) or Eq. (\ref{eq:rel2}) exists in doubly heavy baryons, and to find out how it affects model building as in Refs. \cite{Godfrey:1991,Cahn:2003,Close:2005,Ebert:2009ua}, in which the authors included spin-orbit and tensor forces to obtain a relation like in Eq. (\ref{eq:rel}). {Looking at Table \ref{table2}, one notices that there is one-to-two correspondence between quantum numbers $j_\ell$ and $k$.
% However, in the case of doubly heavy baryons, correspondence becomes one-to-two between them. See Table \ref{tab:Symbol2}.}

%%%%%%%%%%%%%%%%%%%%%%%%%%%%%%%%%%%%%%%%%%%%%%%%%%%%%%%%%%%%%%%%%%%%%%%%%%%%%%%%
\section{Doubly Heavy Baryon}\label{dhbaryons}

We consider the following situation drawn in Fig. \ref{figLight}, where light degrees of freedom can be described by the angular momentum $\vec j_\lambda=\vec l_\lambda + \vec s_q$. The standard $\rho$-$\lambda$ configurations is used for a doubly heavy baryon, in which $\rho$ expresses a length between two heavy quarks in a heavy diquark and $\lambda$ connects a heavy diquark and a light quark. A light quark is attached to one end while an anti-heavy quark $\bar Q$ or a heavy diquark $QQ$ to the other end. The former corresponds to a heavy-light meson and the latter to a doubly-heavy baryon. Irrespective of a heavy object, $\bar Q$ or $QQ$, at one end, light degrees of freedom, i.e., $j_\lambda$, can classify heavy-light systems because it is conserved in the heavy quark limit, $\left[H_{\bar Qq/QQq}, j_\ell\right]=0$.
\begin{figure}[tbp]
\begin{center}
%\scalebox{0.2}{\includegraphics{light.eps}}
\scalebox{0.4}{\includegraphics{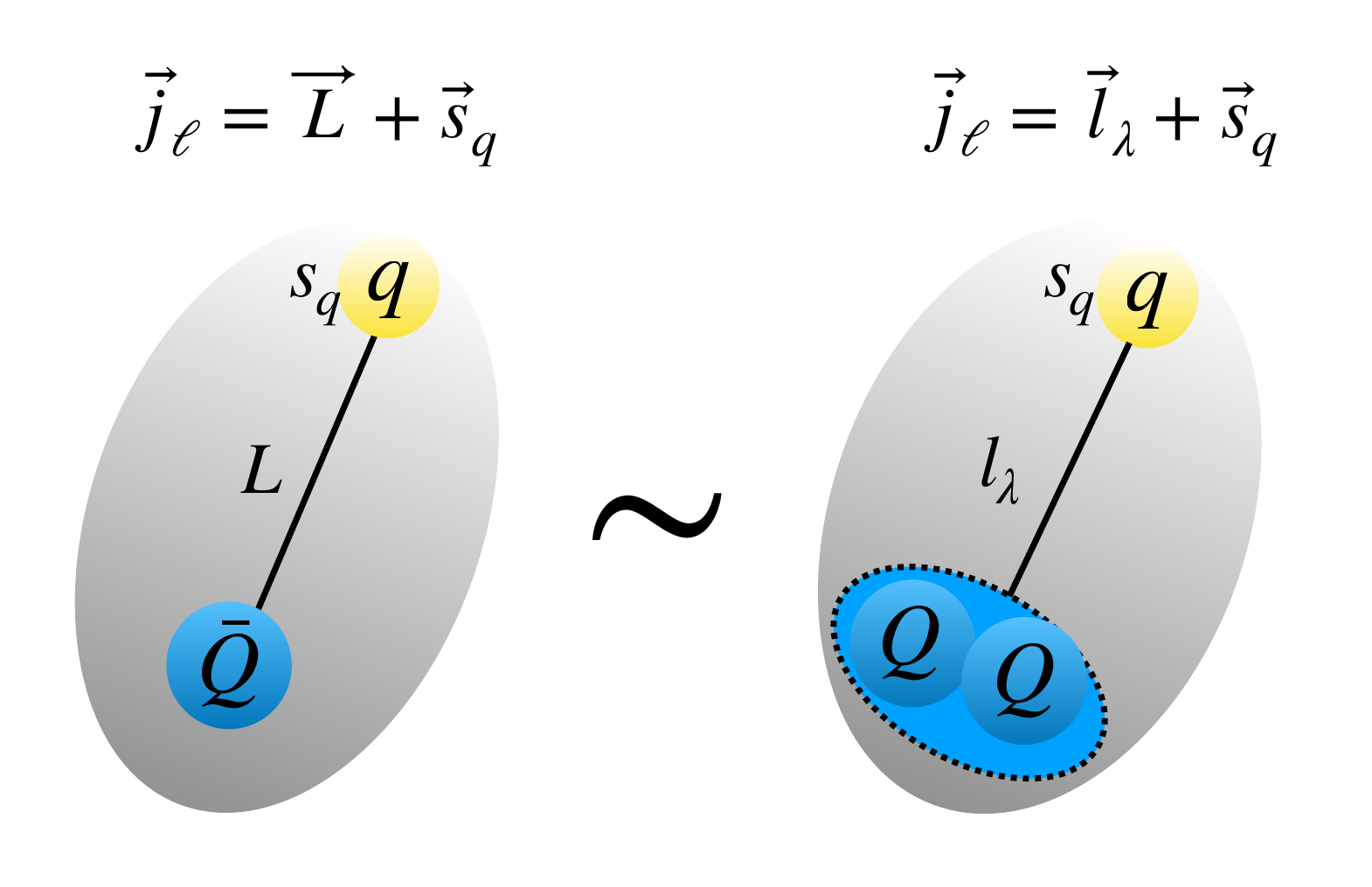}}
\caption{Light degrees of freedom. \label{figLight}}
 \end{center}
\end{figure}

Let us find the heavy-quark symmetric Hamiltonian for a doubly heavy baryon, which can be derived using the same method used in Ref. \cite{Matsuki:1997da}, and is given by \cite{Matsuki:2019xrb},
\begin{eqnarray}
  H_{QQq} = \vec p_\lambda\cdot\vec\alpha_q + \beta_q m_q + 2S(r_\lambda) +V(r_\lambda),
  \label{eq:HQQq}
\end{eqnarray}
where $S(r)$ and $V(r)$ are the same as in Eq. (\ref{eq:HQq}), and $r_\lambda$ is $\sqrt{6}/2$ times the relative distance ($\lambda$) between a light quark and a center of a heavy diquark. This equation can be physically and intuitively understandable, which expresses an interaction between one light quark and two heavy quarks represented by a heavy diquark because there are two color static sources, and a heavy diquark behaves as $\bar 3$ under color $SU(3)$. Since Eq. (\ref{eq:HQQq}) has the same form as Eq. (\ref{eq:HQq}) except for interaction terms, we have the same conserved quantum operator as Eq. (\ref{eq:K}), i.e., $\left[H_{QQq},K\right]=0$, where replacement of $L$ with $l_\lambda$ is tacitly understood.
{Another important point that one should keep in mind is that because the same expression for the Hamiltonian (except for $2S(r)$) is used, we expect that only the same nonrelativistic interaction term, i.e., $\vec l_\lambda\cdot\vec\sigma_q$, keeps the system heavy quark symmetric. This is supported by looking at eigenvalues of $K$ that have the same expression both for heavy-light mesons and doubly heavy baryons, i.e., $k=\pm(j_\ell+1/2)$.\cite{Matsuki:2004zu}}

Now, we have to find a nonrelativistic quantum number instead of $^{2S+1}L_J$ to express heavy quark symmetric states for doubly heavy baryons. Simple extension of $^{2S+1}L_J$ to doubly heavy baryons is given by $^{2S+1}l_\lambda$, where the following replacement is understood,
\begin{eqnarray}
  \vec S \to \vec S= \vec j_\rho+\vec s_q = \vec s_\rho+\vec L_\rho+\vec s_q = \vec s_{Q1}+\vec s_{Q2}+\vec L_\rho+\vec s_q, \quad L\to l_\lambda,
\end{eqnarray}
because $\vec S=\vec s_{\bar Q}+\vec s_q$ for a heavy-light meson corresponds to $\vec S= \vec j_\rho+\vec s_q$ for a doubly heavy baryon.
That is, we adopt $^{2S+1}l_\lambda$ to classify nonrelativistic doubly heavy baryons, where we omit total angular momentum $J$ for simplicity.
Here, we consider only the case that two heavy quarks have the same flavor. In the case that one heavy quark is $c$ and another is $b$, the following discussion cannot be applied.
A wave function of a heavy diquark should be totally antisymmetric or odd in angular momentum, spin, flavor, and color. Hence, angular momentum and spin must be symmetric in exchange of two heavy quarks because two heavy quarks are even in flavor and odd in color. Namely, when $L_\rho$ is even, $s_\rho=s_{Q1}+s_{Q2}=1$, and when $L_\rho$ is odd, $s_\rho=0$. For instance, the total spin is given by $S=1\otimes 1/2=1/2\oplus 3/2$ where $L_\rho=0$ and $s_\rho=1$ for a state with $(N_\rho L_\rho n_\lambda l_\lambda)=(1S1p)$. Here we have used a notation of Ref. \cite{Ebert:2002ig}. Hence $2S+1=2$ or 4.
{There are only five states (spin:1/2,3/2,1/2,3/2,5/2) in the (1S1p) multiplet. However, when $j_\ell \ge 3/2$ with $L_\rho=0$, there are six states in a spin multiplet as seen in Table \ref{tab:Symbol2}.}

Let us find possible mixing states for $N_\rho=n_\lambda=1$, $L_\rho=0$, $s_\rho=1$, and $l_\lambda=1$, i.e., the lowest excited $(N_\rho L_\rho n_\lambda l_\lambda)=(1S1p)$ states. In this case, there are two possible total angular mommenta, i.e., $J=1/2$ and $3/2$, where both of parities are ``$(-)^{L_\rho+l_\lambda}=-$''. As you can see from Table \ref{tab:Symbol2} and using the notation $(N_\rho L_\rho n_\lambda)J^P$ for a state, the $(1S1p)1/2^-$ and $(1S1p)3/2^-$ with $j_\ell=1/2$ or $k=1$ mix with the $(1S1p)1/2^-$ and $(1S1p)3/2^-$ with $j_\ell=3/2$ or $k=-2$, respectively. There is one more state with spin $1/2^-$ coming from $(1P1s)$, which, however, cannot mix with $(1S1p)$ states because matrix elements of physical quantities between $(1S1p)1/2^-$ and $(1P1s)1/2^-$ vanish, which can be shown using $6J$ symbols. For $l_\lambda=2$ states, the $(1S1d)3/2^+$ and $(1S1d)5/2^+$ with $j_\ell=3/2$ or $k=2$ mix with the $(1S1d)3/2^+$ and $(1S1d)5/2^+$ with $j_\ell=5/2$ or $k=-3$, respectively. 
That is, there occurs mixing between $^2l_\lambda$ and $^4l_\lambda$.
In general, for any $l_\lambda$, the $(1S1l_\lambda)(l_\lambda-1/2)^{(-1)^{l_\lambda}}$ and $(1S1l_\lambda)(l_\lambda+1/2)^{(-1)^{l_\lambda}}$ with $j_\ell=l_\lambda-1/2$ or $k=l_\lambda$ mix with the $(1S1l_\lambda)(l_\lambda-1/2)^{(-1)^{l_\lambda}}$ and $(1S1l_\lambda)(l_\lambda+1/2)^{(-1)^{l_\lambda}}$ with $j_\ell=l_\lambda+1/2$ or $k=-l_\lambda-1$, respectively, as seen in Table \ref{tab:Symbol2}.

\begin{table}[htpb]
\centering \caption{The quantum number of the first few states together with those for any $l_\lambda$ with $L_\rho=0$ and $l_\lambda>=2$ In this case that we consider, all the states have $s_\rho=1$ and $L_\rho=0$, and hence $j_\rho=L_\rho\otimes s_\rho=1$. Here, $s_q$ is a light quark spin.\label{tab:Symbol2}}
\begin{tabular}{ccccccc}
\toprule[1pt]
%State        &~& &~&  &~&  \\
$\left(N_\rho L_\rho n_\lambda l_\lambda\right)$  &~& $J^P$ &~& $j_\ell=l_\lambda\otimes s_q$ &~& $k$\\
\midrule[0.6pt] %
$(1S1s)$ &~& $1/2^+$ &~& 1/2 &~& -1\\
$(1S1s)$ &~& $3/2^+$ &~& 1/2 &~& -1\\
\hline 
$(1S1p)$ &~& $1/2^-$ &~& 1/2 &~&  1\\
$(1S1p)$ &~& $3/2^-$ &~& 1/2 &~&  1\\
\hline 
$(1S1p)$ &~& $1/2^{\prime -}$ &~& 3/2 &~& -2\\
$(1S1p)$ &~& $3/2^{\prime -}$ &~& 3/2 &~& -2\\
$(1S1p)$ &~& $5/2^-$ &~& 3/2 &~& -2\\
\hline 
$(1S1d)$ &~& $1/2^+$ &~& 3/2 &~&  2\\
$(1S1d)$ &~& $3/2^+$ &~& 3/2 &~&  2\\
$(1S1d)$ &~& $5/2^+$ &~& 3/2 &~&  2\\
\hline 
$(1S1d)$ &~& $3/2^{\prime +}$ &~& 5/2 &~& -3\\
$(1S1d)$ &~& $5/2^{\prime +}$ &~& 5/2 &~& -3\\
$(1S1d)$ &~& $7/2^+$ &~& 5/2 &~& -3\\
%%%%%%%%%%%%
\hline 
$\vdots$ &~& $\vdots$ &~& $\vdots$ &~& $\vdots$\\
%%%%%%%%%%%%
\hline 
$(1S1l_\lambda)$ &~& $(l_\lambda-3/2)^{(-)^{l_\lambda}}$ &~& $l_\lambda-1/2$ &~& $l_\lambda$\\
$(1S1l_\lambda)$ &~& $(l_\lambda-1/2)^{(-)^{l_\lambda}}$ &~& $l_\lambda-1/2$ &~&  $l_\lambda$\\
$(1S1l_\lambda)$ &~& $(l_\lambda+1/2)^{(-)^{l_\lambda}}$ &~& $l_\lambda-1/2$ &~&  $l_\lambda$\\
\hline 
$(1S1l_\lambda)$ &~& ${(l_\lambda-1/2)^{(-)^{l_\lambda}}}'$ &~& $l_\lambda+1/2$ &~& $-l_\lambda-1$\\
$(1S1l_\lambda)$ &~& ${(l_\lambda+1/2)^{(-)^{l_\lambda}}}'$ &~& $l_\lambda+1/2$ &~& $-l_\lambda-1$\\
$(1S1l_\lambda)$ &~& ${(l_\lambda+3/2)^{(-)^{l_\lambda}}}$  &~& $l_\lambda+1/2$ &~& $-l_\lambda-1$\\
\bottomrule[1pt]
\end{tabular}
\end{table}
{Looking at Table \ref{tab:Symbol2}, one notices that there is one-to-two correspondence again between $j_\ell$ and $k$ as in the case of heavy-light mesons in Table \ref{table2}. One also notices parity for a doubly heavy barons can be given in the following equation: $P = \frac{k}{|k|}(-)^k$. which can be compared with that for heavy-light mesons \cite{Matsuki:2004zu}, $P = \frac{k}{|k|}(-)^{k+1}$.
}

Using the $6J$ symbol, we can calculate extension of Eqs. (\ref{eq:rel2}, \ref{eq:rel3}) to any $l_\lambda$, which can be given by
\begin{eqnarray}
%
%%%% J=l_\lambda+-/2
%
\left( {\begin{array}{*{20}{c}}
   {\left| {{J=l_\lambda-1/2 },j_\lambda = l_\lambda+1/2} \right\rangle }  \\ % upper and lower are interchanged
   {\left| {{J=l_\lambda-1/2 },j_\lambda = l_\lambda-1/2} \right\rangle }  \\
 \end{array} } \right) 
 &=&
 \frac{1}{\sqrt{6l_\lambda+3}}\left( {\begin{array}{*{20}{c}}
   {2\sqrt{l_\lambda+1} } & { \sqrt{2l_\lambda-1} }  \\
   { - \sqrt{2l_\lambda-1} } & { 2\sqrt{l_\lambda+1} }  \\

 \end{array} } \right)\left( {\begin{array}{*{20}{c}}
   {\left| {J=l_\lambda-1/2, S=1/2} \right\rangle }  \\
   {\left| {J=l_\lambda-1/2, S=3/2} \right\rangle }  \\
 \end{array} } \right), \label{DHmix1}\\
%
%%%% J=l_\lambda+1/2
%
\left( {\begin{array}{*{20}{c}}
   {\left| {{J=l_\lambda+1/2 },j_\lambda = l_\lambda+1/2} \right\rangle }  \\ % upper and lower are interchanged
   {\left| {{J=l_\lambda+1/2 },j_\lambda = l_\lambda-1/2} \right\rangle }  \\
 \end{array} } \right) 
 &=&
 \frac{1}{\sqrt{6l_\lambda+3}}\left( {\begin{array}{*{20}{c}}
   {\sqrt{2l_\lambda+3} } & { 2\sqrt{l_\lambda} }  \\
   { - 2\sqrt{l_\lambda} } & { \sqrt{2l_\lambda+3} }  \\

 \end{array} } \right)\left( {\begin{array}{*{20}{c}}
   {\left| {J=l_\lambda+1/2, S=1/2} \right\rangle }  \\
   {\left| {J=l_\lambda+1/2, S=3/2} \right\rangle }  \\
 \end{array} } \right), \label{DHmix2}
 \label{eq:rel4}
\end{eqnarray}
where magic mixing angles $\theta_m$ are given by
\begin{eqnarray}
  \tan\theta_m &=& \frac{\sqrt{2l_\lambda-1}}{2\sqrt{l_\lambda+1}} \quad({\rm for}~~J=l_\lambda-1/2), \label{magic1}\\
  \tan\theta_m &=& \frac{2\sqrt{l_\lambda}}{\sqrt{2l_\lambda+3}} \quad({\rm for}~~J=l_\lambda+1/2), \label{magic2}
\end{eqnarray}
and parities are given by $(-)^{L_\rho+l_\lambda}=(-)^{l_\lambda}$. For instance, for $l_\lambda=1$, $\tan\theta_m=1/(2\sqrt{2})$ for $J^P=1/2^-$ and $2/\sqrt{5}$ for $J^P=3/2^-$ corresponding to $\theta_m=15.9^\circ$ and $41.8^\circ$, respectively.

Let us consider how this relation affects the results obtained so far in other papers. For instantce, let us consider the results obtained in Ref.~\cite{Lu:2017meb}, in which quantum numbers are carefully taken care of so that our $S$ for a doubly heavy baryon is denoted as $S_T$. 
In order to obtain heavy quark symmetric states for doubly heavy baryons, which are considered to be closer to physical states, 
the authors seemed to neglect some interaction terms to keep heavy quark symmetry intact and to succeed in classifying the states in terms of $j_\ell$.
The authors may need to further investigate the mixing mechanism by including some appropriate interactions.
%they need to include interactions like those mentioned below in Eq. (\ref{eq:Hdh}). 
Another paper Ref. \cite{Ebert:2002ig} starts from heavy quark symmetric interactions so that they could classify the states in terms of $j_\ell=l_\lambda + s_q$ but did not obtain the relations like in Eqs. (\ref{DHmix1}, \ref{DHmix2}) derived in this paper.

We should mention the case where $L_\rho\ne 0$. Since tthe case $L_\rho=1$ is very interesting, we describe it more details here. Since $s_\rho=0$ in this case, we obtain $j_\rho=s_\rho\otimes L_\rho=1$ and we obtain the same table as Table \ref{tab:Symbol2} with the opposite parity for each baryon. Magic mixing angles are also given by the same equations Eqs. (\ref{DHmix1}-\ref{magic2}) between appropriate states. Because other cases where $L_\rho\ge 2$ become complicated, we give it to a furure work.

%%%%%%%%%%%%%%%%%%%%%%%%%%%%%%%%%%%%%%%%%%%%%%%%%%%
\section{Summary and discussions}\label{summary}
Defining such a magic mixing angle that one can obtain heavy-quark symmetric states by rotating nonrelativistic states with definite $^{2S+1}l_\lambda$, we derive mixing matrices in Eqs. (\ref{DHmix1}, \ref{DHmix2}) using the $6J$ symbol. Since the nature prefers the heavy quark symmetry, the heavy-quark symmetric states are considered to be physical ones. Although we have succeeded in obtaining magic mixing angles in the case of heavy-light mesons in Ref. \cite{Matsuki:1997da} from the heavy quark symmetric Hamiltonian, we could not derive such a Hamiltonian for doubly heavy baryons in this paper. See one trial in Ref. \cite{Matsuki:2019xrb}. However, following the derivation in the case of heavy-light mesons and using the $6J$ symbol, we have succeeded in obtaining magic mixing angles for any $l_\lambda$ in Eqs. (\ref{magic1}, \ref{magic2}) for $J=l_\lambda\pm1/2$ after deriving the relations between two states with the same quantum number $(N_\rho L_\rho n_\lambda l_\lambda)J^P$ but with the different $j_\ell$ or $k$ quantum number. {A recent paper written by one of the authors \cite{He:2021iwx} studied strong decays of doubly bottom baryons. They related the $J-J$ coupling to the $L-S$ coupling and hence they tacitly used the magic mixing angles between the states with the same $J^P$, whose results, of course, depend on the magic mixing angles we have studied in this paper.
In a paper \cite{Xiao:2017dly}, they studied strong decays of low-lying doubly charmed baryons. They noticed that the $\lambda$ mode excited states should be classified in $^2p/^4p$ quantum numbers for $(1S1p)1/2^-$, $(1S1p)3/2^-$, etc. with $j_\ell=1/2,~3/2$ corresponding to $k=1,~-2$, respectively. However, they did not treat them as mixed states studied in this paper. So there is some confusion how to calculate strong decays (e.g., one chiral particle decays) of $\lambda$ mode excited states, i.e., $\Xi_{cc/bb}$ and $\Omega_{cc/bb}$.
}  

%%%%%%%%%%%%%%%%%%%%%%%
%Here, we have used $P= (-1)^{|k| + 1}\;{k}/{|k|}$ with $k=L_\lambda$.
%%%%%%%%%%%%%%%%%%%%%%%

A standard way to calculate a spectrum of doubly heavy baryons is i) to first calculate heavy diquark mass, and then, ii) regarding the whole system as a heavy-light system, to apply the potential model like a heavy-light meson. In this situation, especially in the second step,  people normally obtain the eigenstates with definite quantum number $^{2S+1}l_\lambda$ since the interactions for a heavy-light system include the spin-orbit or spin-spin interaction like $\vec l_\lambda\cdot\vec S$ or $\vec s_\rho\cdot\vec s_q$, which breaks the heavy quark symmetry (HQS) as shown in, e.g., Ref. \cite{Matsuki:2010zy}. 
{Another standard way might be solving the three body system in the quark model like Ref. \cite{Capstick:1986bm}.}
To obtain the heavy quark symmetric spectrum  for doubly heavy baryons, one needs to explicitly exclude such terms that break the HQS or include only the term, $\vec\sigma_q\cdot\vec l_\lambda$, that keeps the HQS. The best way is to first obtain the spectrum of doubly heavy baryons using the standard way with the potential model, and rotate the obtained wave functions as well as masses with mixing matrices Eqs. (\ref{DHmix1}, \ref{DHmix2}) although there appear off-diagonal elements in the mass matrix that should be neglected as an approximation.

Ref. \cite{Ebert:2002ig} treated doubly heavy baryons very carefully. They start from the heavy quark symmetric Hamiltonian in the limit of $M_d\to\infty$ with $M_d$ heavy diquark mass, and then, include $1/M_d$ corrections to improve the results. This operation causes mixing between heavy quark symmetric states. For instance, they explicitly showed mixing between $(1S2p)1/2^-$ with $j_\ell=1/2$ or $k=1$ mix with the $(1S2p)1/2^-$ with $j_\ell=3/2$ or $k=-2$, which is nothing but breaking of the HQS. This is a different problem we have considered in this article. They did not consider the case that the heavy quark symmetric states consist of nonrelativistic states with definite $^4p$ and/or $^2p$ quantum numbers, and hence did not derive the realtions we obtained in Eqs. (\ref{DHmix1}, \ref{DHmix2}).

One of the other novel things to be mentioned is that there are other possible admixtures other than those shown in Table \ref{tab:Symbol2} that respect the HQS. For instance, one can imagine states with higher $L_\rho \ge 2$, e.g., $(L_\rho=3, 5, \cdots, s_\rho=0)$ and $(L_\rho=2, 4, 6, \cdots, s_\rho=1)$. At present, to derive equations corresponding to Eqs. (\ref{DHmix1}-\ref{magic2}),we need to study case by case and leave this as a future work, which is not a difficult task if we appropriately use the $6J$ symbol. Another thing to be mentioned is that to classify a doubly heavy baryon spectrum, the quantum number $K$ is preferable compared with $j_\ell$ because as seen in Table \ref{tab:Symbol2}, even the states with the same $j_\ell$ have different $k$ values. {The eigenvalue $k$ gives the same expression for a relation with $j_\ell$, i.e, $k=\pm(j_\ell+1/2)$, but gives different expression for parity, $P = \frac{k}{|k|}(-)^{k+1}$ and $\frac{k}{|k|}(-)^k$ for heavy-light mesons and doubly heavy baryons, respectively.}

\section*{Acknowledgement}

D.-Y. Chen is partly supported by the National Natural Science Foundation of China under the Grant Nos. 11775050 and 11675228. X. Liu is  partly  supported  by  the  China  National Funds  for  Distinguished  Young  Scientists  under  Grant No.11825503 and the National Program for Support of Top-notchYoung Professionals. Q.-F. L\"u is partly supported by the National Natural Science Foundation of China under Grant Nos. 11705056 and U1832173, and by the State Scholarship Fund of China Scholarship Council under Grant No. 202006725011.

\end{document}